\documentclass[a4paper]{panl}
\usepackage{cite}
\usepackage{wrapfig}
\usepackage{graphicx}
\usepackage{amssymb}
\usepackage{amsfonts}
\usepackage{amsmath,amsthm}
\usepackage{longtable}
\usepackage{rotating}
\usepackage{lscape}
\usepackage{epsfig}
\usepackage{multirow}
\usepackage{csvsimple}
\usepackage[english]{babel}

\originalTeX
\graphicspath{{./Figures/}}             
\usepackage{graphicx,url,color}

\newcommand{\beq}{\begin{equation}}
\newcommand{\eeq}{\end{equation}}
\newcommand{\bea}{\begin{eqnarray}}
\newcommand{\eea}{\end{eqnarray}}

\newcommand{\eq}{\begin{equation}}
\newcommand{\en}{\end{equation}}
\newcommand{\eqa}{\begin{eqnarray}}
\newcommand{\ena}{\end{eqnarray}}

\begin{document}
\title{ Gribov copy effects in the maximal Abelian gauge}

\maketitle
\authors{I.\,Kudrov$^{a,}$\footnote{E-mail: Ilya.Kudrov@ihep.ru},
V.\,Bornyakov$^{a,b,c,}$\footnote{E-mail: Vitaly.Bornyakov@ihep.ru}}
\setcounter{footnote}{0}
\from{$^{a}$\, NRC ``Kurchatov Institute'' - IHEP, Protvino, 142281 Russia }
\from{$^{b}$\, Pacific Quantum Center, Far Eastern Federal University, 690950 Vladivostok, Russia}
\from{$^{c}$\, KCTEP, NRC “Kurchatov Institute”,  Moscow, Russia}




\begin{abstract}
We study effects of Gribov copies in the Maximal Abelian gauge in $SU(3)$ lattice gluodynamics. 
We confirm earlier results that with effective gauge fixing algorithm one finds Gribov copies of the maximal Abelian gauge which produce about 90\% of the nonabelian string tension after Abelian projection. At the same time using much less effective relaxation algorithm one finds Gribov copies with nice Abelian dominance for the string tension.

\end{abstract}

\vspace*{6pt}




\section{Introduction}
\label{section1}

In this paper, we numerically study lattice gluodynamics with gauge group $SU(3)$ 
in the maximal Abelian (MA) gauge introduced in \cite{thooft2} and formulated for lattice regularization in the works of \cite{Kronfeld:1987vd,Brandstater:1991sn}. The gauge fixing functional is defined as follows
\beq
F =  \frac{1}{V} \int d^4 x  \sum_{\mu, a \neq 3,8} A^a_\mu(x) A^a_\mu(x),
\label{functional}
\eeq
where $A^a_\mu(x)$ is the gauge field. The functional (\ref{functional}) is invariant under Abelian gauge transformations $g(x) \in U(1) \times U(1)$.
In lattice regularization, minimizing the gauge functional (\ref{functional}) is equivalent to maximizing the functional
\begin{equation}
F_{lat} = 
\frac{1}{12\,V}
 \sum_{x,\mu}\left[ |U^{(11)}_\mu(x)|^2 +|U^{(22)}_\mu(x)|^2 
            +|U^{(33)}_\mu(x)|^2 \right]
\label{functional2}
\end{equation}

The functional (\ref{functional}) has numerous local minima corresponding to the Gribov copies discovered by Gribov for the Coulomb gauge in~\cite{Gribov:1977wm}. In the framework of perturbation theory, this problem does not appear and quantization can be successfully performed using the Faddeev-Popov method. However, in the non-perturbative domain, the Faddeev-Popov method fails because there are many gauge-equivalent configurations, called Gribov copies, satisfying the given gauge condition. 
Gribov's result was generalized to other gauges in \cite{Singer:1978dk}. 

Non-perturbatively, a gauge fixing can be defined as follows \cite{Zwanziger:1990tn, Parrinello:1990pm}
\beq
\langle
{\cal{O}}\rangle = \frac{1}{Z(\lambda)} \int DA~e^{-S(A)}~I^{-1}(A) \int
Dg~e^{-\lambda F(A^g)} {\cal{O}}(A^g) \,, 
\label{pjlz1}
\eeq
where $\lambda$ - gauge parameter, $\cal{O}$  - some observable,
\begin{equation}
    I(A) = \int Dg ~e^{-\lambda F(A^g)}\,.
\end{equation}

The limit $\lambda \to \infty$ corresponds to the restriction of the integration in the functional integral to the fundamental modular domain. Such a way of solving the Gribov copies problem was proposed in~\cite{Zwanziger:1993dh}. 
Lattice regularization allows us to study the effects of Gribov copies by numerical methods. Strong Gribov copy effects, i.e., a strong dependence of gauge non-invariant observables on the choice of Gribov copies \cite{Bali:1996zs}, have been found in MA gauge. It is practically impossible to find global minima of the gauge functional numerically, but it is natural to assume that by generating many such minima and taking the minimal of them, we approach the global minimum. Such a practical approach to reducing the effect of Gribov copies was proposed in~\cite{Bali:1996zs}, where the MA gauge was studied in lattice $SU(2)$-gluodynamics. This approach was then used in studies of MA gauge in both gluodynamics \cite{Bali:1996dm} and QCD \cite{DIK:2003alb}, as well as in studies of other gauges \cite{Bornyakov:2000ig,Bornyakov:2008yx,Bornyakov:2010nc}.

The most effective algorithm for fixing the gauge, if the search for the global minimum of the gauge functional is required, is the simulated annealing algorithm. Another, less effective, but often used in practice algorithm is local relaxation (minimization). Both algorithms are briefly described in the next section. In the case of MA gauge, these algorithms give significantly different results both for the gauge functional and for physically interesting quantities such as the Abelian string tension $\sigma_{ab}$. In the case of $SU(2)$ gluodynamics, it has been shown that the relaxation algorithm can give for $\sigma_{ab}$ a value equal to or even greater than the value of the non-Abelian string tension $\sigma$, while by using for a gauge fixing procedure the simulated annealing algorithm one gets $\sigma_{ab} < \sigma$ \cite{Bali:1996dm}, at least for a finite lattice spacing. In \cite{bm} it was shown that in the case of $SU(2)$ gluodynamics in the continuum limit, the above optimal gauge fixing procedure leads to $\sigma_{ab} \approx \sigma$. In the case of $SU(3)$ gluodynamics the situation is less certain. The effects of Gribov copies were investigated in \cite{DIK:2003alb} using a simulated annealing algorithm. A strong dependence of $\sigma_{ab}$ on the choice of Gribov copy was demonstrated; for a lattice spacing $a \approx 0.1$~Fm, $\sigma_{ab}/\sigma \approx 0.83$ was obtained. Later, the authors of Ref. \cite{Sakumichi:2014xpa} concluded that $\sigma_{ab} \approx \sigma$ for lattice spacing $a \approx 0.1$~Fm or less, and this result critically depends on the physical size of the lattice. Below we present results that resolve the contradiction between the conclusions of \cite{DIK:2003alb} and \cite{Sakumichi:2014xpa}.

\section{Simulation details}
Table \ref{table1} summarizes the parameters (lattice spacing $a$, lattice size, number of configurations) of the lattices used in this work. The Wilson lattice action was used to generate the lattice gauge field configurations. The Sommer parameter \cite{Sommer:1993ce} was used to determine the lattice spacing in physical units, with values of $r_0/a$ taken from \cite{Necco:2001xg}. In gluodynamics it is common to use the value of this parameter $r_0=0.5$ Fm.

\begin{table}[hbtp]
\begin{center}
\begin{tabular}{|cccccc|}  \hline
$\beta$       &  $r_{0}/a$   & $ a$, Фм    &  $L/a$& $L$, Фм & $N_{conf}$      \\
\hline
$6.0$ & 5.37 & 0.093 & 16 & 1.49 & 1600 \\
$6.0$ & 5.37 & 0.093 & 24 & 2.24 & 5000 \\
$6.0$ & 5.37 & 0.093 & 32 & 2.98 & 4000  \\
$6.1$ & 6.34 & 0.079 & 28 & 2.21 & 5000 \\
$6.2$ & 7.38 & 0.068 & 32 & 2.17 & 3800 \\
$6.3$ & 8.51 & 0.059 & 36 & 2.11 & 2100 \\
\hline
\end{tabular}
\end{center}
\caption{The parameters of lattices used in this work. }
\label{table1}
\end{table}

\begin{table}[hbtp]
\begin{center}
\begin{tabular}{|cccccc|}  \hline
$\beta$ & $L/a$ & $F^{n_{copy}=1}_{RO}$ & $F^{n_{copy}=max}_{RO}$ & $F^{n_{copy}=1}_{SA}$ & $F^{n_{copy}=max}_{SA}$ \\
\hline
$6.0$ & 16 & 0.73216 & 0.73317 & 0.73407 & 0.73431  \\
$6.0$ & 24 & 0.73224 & 0.73272 & 0.73404 & 0.73424  \\
$6.0$ & 32 & 0.73226 & 0.73253 & 0.73403 & 0.73415  \\
$6.1$ & 28 & 0.74216 & 0.74255 & 0.74310 & 0.74349  \\
$6.2$ & 32 & 0.75098 & 0.75131 & 0.75169 & 0.75204  \\
$6.3$ & 36 & 0.75894 & 0.75923 & 0.76003 & 0.76010  \\
\hline
\end{tabular}
\end{center}
\caption{The resulting functional values (\ref{functional2}) using relaxation only (RO) and using simulated annealing (SA) for the first gauge copy and for the best gauge copy. The error is in the fifth digit.}
\label{table2}
\end{table}

To fix the MA gauge on each lattice configuration, a gauge transformation was found that maximizes the functional 
(\ref{functional2}). This was done using two algorithms. The first is the relaxation algorithm: at each lattice site, a gauge transformation is found site by site that locally maximizes the functional (\ref{functional2}) until a maximum of this functional is found. The second algorithm, simulated annealing, is a more efficient algorithm that is applied before the relaxation algorithm and it provides a higher probability of obtaining a higher value of the functional after relaxation. At each step of this algorithm, the gauge transformation is updated using the heat bath algorithm. The local gauge transformation $g(x)$ is chosen with probability $\propto e^{F_{local}/T_{SA}}$, where $T_{SA}$ is the effective temperature of this algorithm and $F_{local}$ is the contribution from a cite $x$ to the functional (\ref{functional2}). Initially, a sufficiently large temperature $T_{SA}$ is chosen, then after each update of the gauge transformation, the temperature is gradually decreased down to some minimal value $T_{SA}$. Then the simulated annealing is terminated and the MA gauge is finally fixed by relaxation. In the case of $SU(3)$ gluodynamics, the optimal temperature range $T_{SA} \in (0.01;1.25)$ was chosen. Both  algorithms find a random local maximum of the functional. Repeating the procedure few times one can find a better value of the local maximum (better Gribov copy) and investigate the dependence on gauge copies. The obtained values of the functional are summarized in the table \ref{table2}.

In this work, we used the implementation of the gauge fixing algorithms presented in \cite{Schrock:2012fj}. The corresponding code can be downloaded from\\ \text{https://github.com/havogt/culgt/tree/master}.

\section{Abelian string tension in the case of the simulated annealing algorithm}
As mentioned above, the most efficient algorithm used to fix a gauge defined by extremization of the gauge functional, such as the Landau gauge, Coulomb gauge, MA gauge, and central gauge, is the simulated annealing algorithm. 
In this section, we present our results for the Abelian string tension $\sigma_{ab}$ obtained using this algorithm. 

The Abelian projection performed after fixing the MA gauge, for a lattice gauge field, means decomposing the non-Abelian gauge field $U_\mu(x) \in SU(3)$ into the product of the non-diagonal component $U^{offd}_\mu(x) \in SU(3)/U(1) \times U(1)$ and the diagonal component $U^{Abel}_\mu(x) \in U(1) \times U(1)$ :
\begin{equation}
    U_\mu(x) = U^{offd}_\mu(x) U^{Abel}_\mu(x) \,.
\end{equation}
$U^{Abel}_\mu(x)$ has the form
\beq
\label{uabel}
U^{Abel}_\mu(x) = \mbox{diag} \left(u^{(1)}_\mu(x),u^{(2)}_\mu(x),u^{(3)}_\mu(x)
\right)\,,
\eeq
where 
\beq
\label{ulink}
u^{(a)}_{\mu}(x)=e^{i \theta^{(a)}_{\mu}(x)}\,, 
\eeq
\beq
\label{tlink}
\theta^{(a)}_{\mu}(x) = \arg~(U_\mu(x))_a-\frac{1}{3} \sum_{b=1}^3
\arg(U_\mu(x))_b\,\big|_{\,{\rm mod}\ 2\pi}\,.
\eeq
and 
\beq
\theta^{(a)}_{\mu}(x) \in [-\frac{4}{3}\pi, \frac{4}{3}\pi]\,.
\eeq
This definition of Abelian projection maximizes the expression \\
$|\mbox{Tr} \left( U_\mu^\dagger(x) U^{Abel}_\mu(x) \right) |^2$.

\begin{figure}[hbt]
\vspace*{-1.5cm}
\begin{center}
\includegraphics[width=127mm]{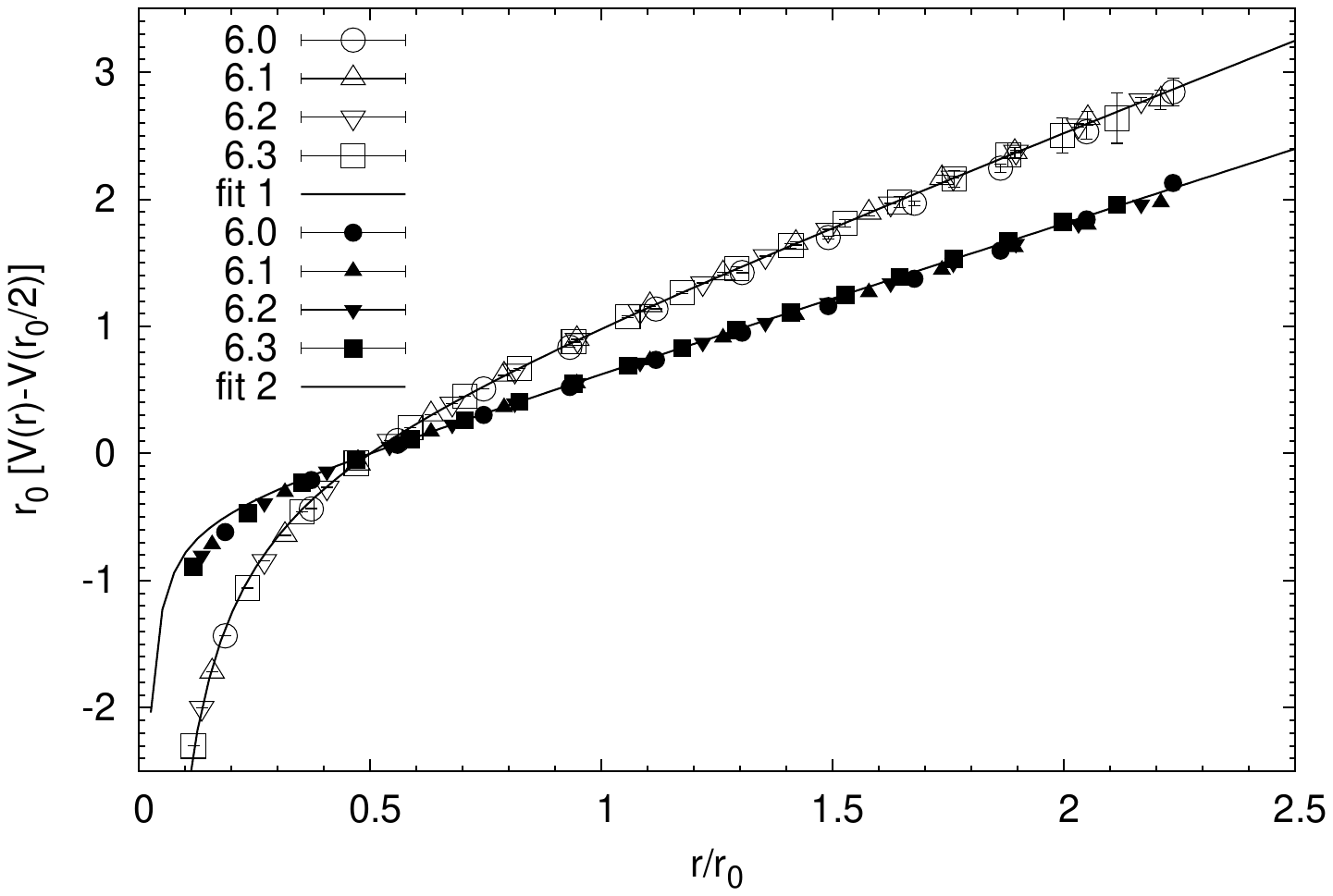}
\vspace{-10mm}
\caption{Non-Abelian (empty symbols) and Abelian (filled symbols) static potentials for four values of lattice spacing. The curves show the results of fits with the Cornell potential in the case $\beta=6.3$.}
\end{center}
\labelf{poten_SA}
\vspace{-5mm}
\end{figure}

It is well known that
in the MA gauge the Abelian string tension $\sigma_{ab}$ computed from the Abelian Wilson loops is quite close to the nonabelian string tension $\sigma$. This observation, confirmed in both gluodynamics and QCD \cite{suzuki1,suzuki2,Bali:1996dm,bm,DIK:2003alb,Sakumichi:2014xpa}, supports the concept of the dominance of Abelian degrees of freedom at large distances (see, e.g., reviews \cite{review1,review2,review3}). 

We performed the standard calculation of the static potential $V(r)$ for four values of lattice spacing given in Table~\ref{table1}. In computation of $V(r)$ we used APE smearing \cite{APE:1987ehd} for links in the spatial directions and one step of the hypercubic blocking \cite{Hasenfratz:2001hp} for links in the time direction of the Wilson loop. We also performed the computation of the Abelian static potential $V_{ab}(r)$ using APE smearing. The results for $V(r)$ and $V_{ab}(r)$, normalized by  $r_0$ are presented in Fig.~\ref{poten_SA}. To exclude the contribution of the divergent source selfenergy, we show the difference $ V(r) - V(r_0/2)$ and similarly for 
$V_{ab}(r)$. The figure shows the fits of the data for $\beta=6.3$ by the Cornell potential
\begin{equation}
    V(r) = V_0 + \alpha/r +\sigma r \,
\end{equation}
for both non-Abelian and Abelian potentials.
The figure shows that the data for $ r_0(V(r) - V(r_0/2))$ obtained for different values of lattice spacing agree well with each other. A similar result is seen for $ r_0(V_{ab}(r) - V_{ab}(r_0/2))$. However, the slopes of the potentials $V(r)$ and $V_{ab}(r)$ are different at large distances. We obtained for the minimal lattice spacing the ratio $\sigma_{ab}/\sigma \approx 0.83(2)$ and similar results for other values of the lattice spacing. This result is in good agreement with the result obtained for $\beta=6.0$ in \cite{DIK:2003alb}. We also find that, in contrast to the case of $SU(2)$ gluodynamics \cite{bm}, the ratio $\sigma_{ab}/\sigma$ does not converge to 1 in the continuum limit. 

The next step is to check for finite volume effects, which are considered in \cite{Sakumichi:2014xpa} as the reason for $\sigma_{ab}/\sigma$ being lower than 1. Figure ~\ref{poten_ab_compar} presents the results for $(V_{ab}(r)$ obtained for $\beta=6.0$ on lattices of size $L=24$ and $L=32$ using the simulated annealing algorithm. The results for $L=24$ are the same as shown in Fig.~\ref{poten_SA}, they are obtained for number of Gribov copies $N_{cop}=4$. In the case of $L=32$, the results obtained for $N_{cop}=20$ are shown. The values of the parameter $N_{cop}$ are chosen so that the value of the functional (\ref{functional2}) turns out to be approximately the same for these two lattices. The values of the functional from Table \ref{table2} for the relaxation algorithm are the same as those given in \cite{Sakumichi:2014xpa}. It can be seen from Fig.~\ref{poten_ab_compar} that changing the lattice size does not affect the slope of the potential $V_{ab}(r)$. This result contradicts the conclusions made in \cite{Sakumichi:2014xpa}.

\begin{figure}[ht]
\vspace*{-1.5cm}
\begin{center}
\includegraphics[width=127mm]{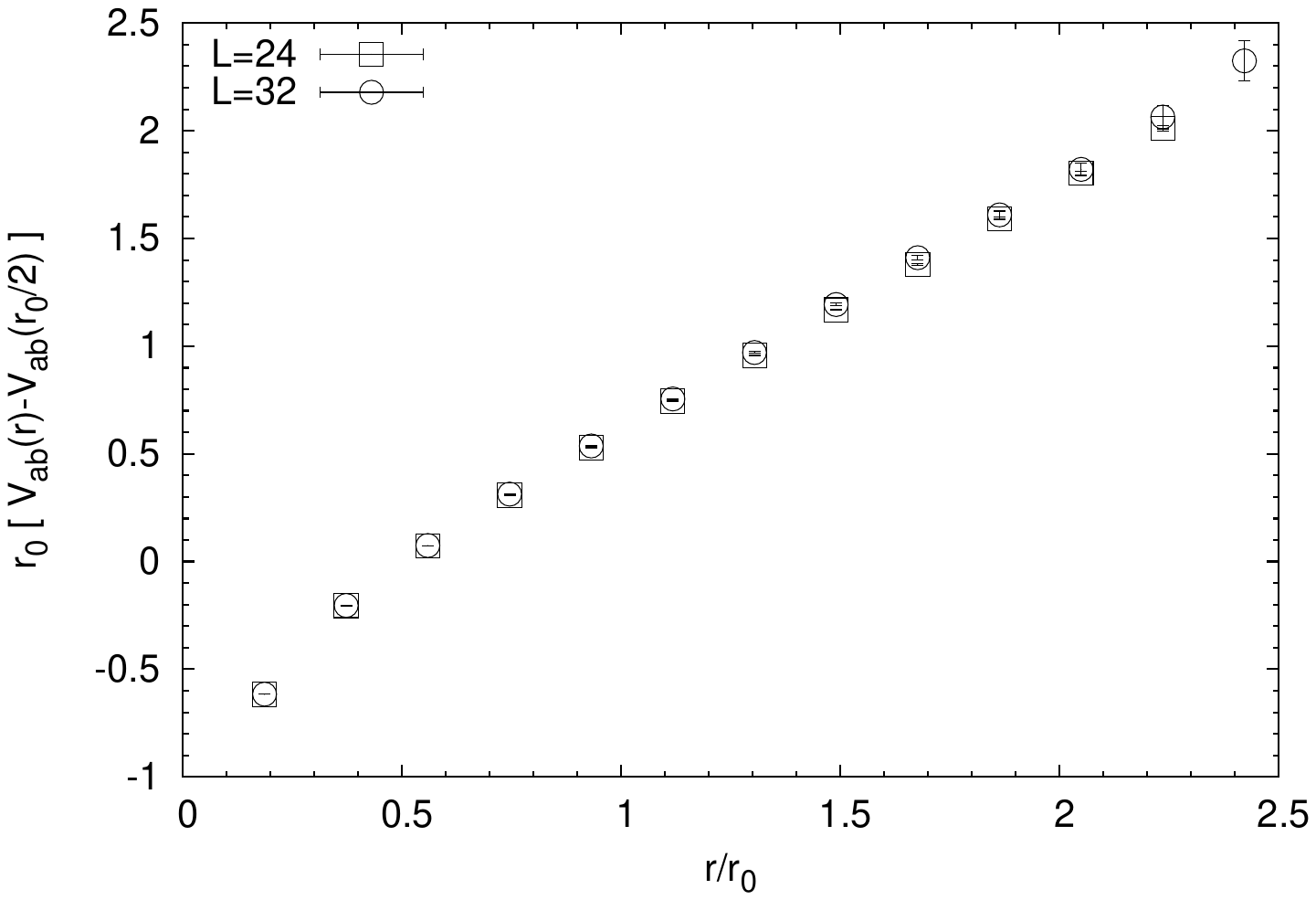}
\vspace{-10mm}
\caption{Comparison of Abelian static potentials on lattices $L=24$ and $L=32$ for $\beta=6.0$.}
\end{center}
\labelf{poten_ab_compar}
\vspace{-5mm}
\end{figure}

\section{Abelian string tension in the case of the relaxation algorithm}
On the lattice gauge field configurations described in the previous section, we also performed MA gauge fixing using the relaxation algorithm, which is less efficient than the simulated annealing algorithm, i.e., it gives lower values for the functional (\ref{functional2}).  Fig.~\ref{poten_RO} shows the results for $V_{ab}(r)$ calculated after such gauge fixing for $N_{cop}=1$. The results for $V(r)$ are represented by the fit curve shown also in Fig.~\ref{poten_SA}. It can be seen from the figure that the data for the Abelian potential calculated for different values of lattice spacing lie on the universal curve as in the previous case, but the slope of the Abelian potential has changed significantly. A good agreement between the slope of the potential $V_{ab}(r)$ and the slope of the potential $V(r)$ can be seen. For the ratio $\sigma_{ab}/\sigma$ we obtained a value of $0.96(3)$. Thus, we confirm earlier results that there exist Gribov copies on which the Abelian string tension is equal to the nonabelian string tension.
\begin{figure}[h]
\vspace*{-1.5cm}
\begin{center}
\includegraphics[width=127mm]{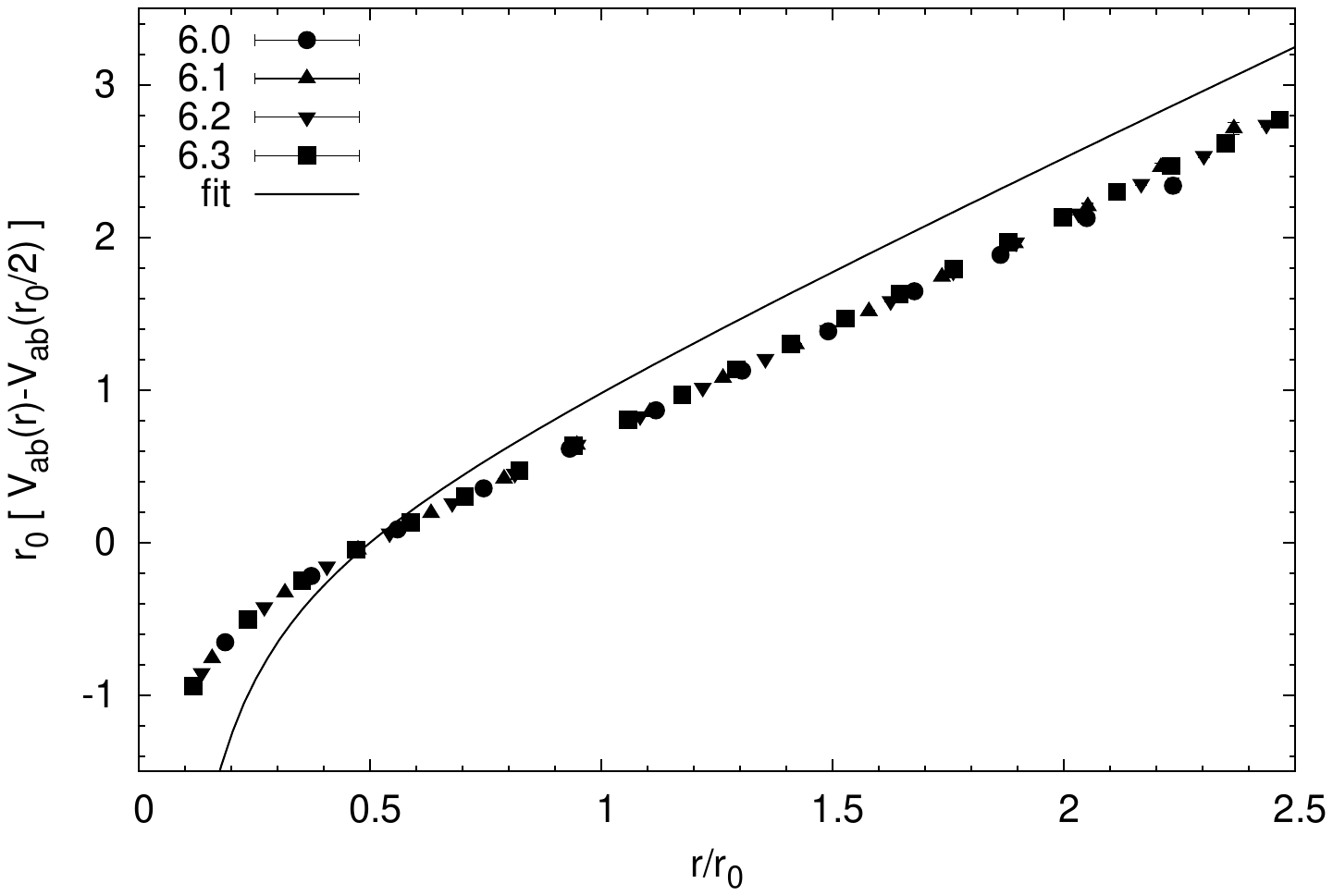}
\vspace{-10mm}
\caption{Abelian static potentials for four values of lattice spacing. The curve shows the fit result by the Cornell potential for the nonabelian potential in the case $\beta=6.3$.}
\end{center}
\labelf{poten_RO}
\vspace{-5mm}
\end{figure}

\section{Conclusions}
In this paper, we have performed a study of Gribov copy effects in MA gauge, which is intensively used to investigate the dual superconductor scenario of the confinement phenomenon \cite{thooft,Mandelstam:1974pi}. Two algorithms were used to fix the MA gauge, which give significantly different values for the gauge functional (\ref{functional2}). An Abelian projection was performed on the gauge field configurations obtained after fixing the gauge, the Abelian static potential $V_{ab}(r)$ was computed, and a comparison with the nonabelian potential $V(r)$ was performed. We shown that on the Gribov copies obtained with the simulated annealing algorithm, $\sigma_{ab}$ is significantly lower than $\sigma$, which is in agreement with a previous result presented in \cite{DIK:2003alb}. It is important to note that in the considered case of $SU(3)$ gluodynamics the ratio $\sigma_{ab}/\sigma$ is independent of the lattice spacing, which is different from the result obtained in \cite{bm}
for $SU(2)$ gluodynamics. 

Comparing the results obtained on lattices of different sizes for a fixed lattice spacing at $\beta=6.0$, we conclude that $\sigma_{ab}$ does not depend on the lattice size but is determined by the value of the gauge functional (\ref{functional2}). This result contradicts the conclusion about the strong influence of finite volume effects on the value of $\sigma_{ab}$ made in \cite{Sakumichi:2014xpa}.

Further, it has been shown that on Gribov copies obtained using the relaxation algorithm, which has a significantly reduced value of the gauge functional (\ref{functional2}), the ratio $\sigma_{ab}/\sigma$ is close to 1 and depends weakly on the lattice spacing. This allows us to conclude that there are Gribov copies on which 'perfect' Abelian dominance can be observed. This 'perfect' Abelian dominance was demonstrated previously in \cite{Sakumichi:2014xpa}, but as noted above, the conclusions made in that paper concerning the finite volume effects were not correct.

Finally, we note that the problem of selection of proper Gribov copies has been discussed intensively for central gauges in \cite{Faber:2001hq,Golubich:2020sqd,Dehghan:2024rly}. In the future, we plan to perform a study of this problem in MA gauge using the ideas formulated in these papers. The choice of Gribov copies on which $\sigma_{ab} \approx \sigma$ is important for us in terms of the decomposition of the gauge field into monopole and monopole-free components, which we investigated in Refs.~\cite{Bornyakov:2021enf,BornyakovVG:2023rci}

\label{sec:acknowledgement}
\section*{Acknowledgement}
The authors are grateful for the following computer resources: the central Linux-cluster of the A.A. Logunov Institute of High Energy Physics of the Kurchatov Institute (Protvino), the Linux-cluster of the KCTEP of the Kurchatov Institute (Moscow), and the “Complex of modeling and data processing of mega-class research facilities” of the Kurchatov Institute (Moscow).

\label{sec:funding}
\section*{Funding}
This work was supported by the Russian Science Foundation, grant 23-12-00072

\bibliographystyle{pepan}
\bibliography{bibliography.bib}

\end{document}